\documentclass[showpacs,aps,amssymb]{aipproc}

\layoutstyle{8x11double}
\usepackage{graphicx}
\usepackage{bm}
\usepackage{amssymb}   
\usepackage{amsmath}
\usepackage{amsfonts}

\def\calO{{\cal O}}
\def\calU{{\cal U}}

\def\hbar{{\bar h}}

\def\lambdabar{{\bar\lambda}}

\def\rhohat{{\hat\rho}}
\def\vhat{{\hat v}}

\def\nn{\nonumber}

\begin{document}
\title{Unparticles and electroweak symmetry breaking}
\author{Jong-Phil Lee}{
address={Korea Institute for Advanced Study, Seoul 130-722, Korea},
email={jplee@kias.re.kr}}

\begin{abstract}
We investigate a scalar potential inspired by the unparticle
sector for the electroweak symmetry breaking. The scalar potential
contains the interaction between the standard model fields and
unparticle sector. It is described by the non-integral power of
fields that originates from the nontrivial scaling dimension of
the unparticle operator. 
It is found that the electroweak symmetry is
broken at tree level when the interaction is turned on. The scale
invariance of unparticle sector is also broken simultaneously,
resulting in a physical Higgs and a new lighter scalar particle.

\end{abstract}

\maketitle
\section{Introduction}
The start-up of the large hadron collider (LHC) at CERN will shed lights on
a long standing puzzle in particle physics,
the secret of electroweak symmetry breaking (EWSB).
A hidden sector can be a good answer for EWSB.
In a minimal extension, the hidden sector scalar couples to the 
standard model (SM) scalar field
in a scale-invariant way \cite{Meissner,Espinosa,CNW,FKV}.
Quantum loop corrections break the scale invariance,
and the scalar field achieves the vacuum expectation value (VEV) through
the Coleman-Weinberg (CW) mechanism \cite{CW}.
\par
Recently the hidden sector has received much attention with the possibility
of the existence of unparticles \cite{Georgi}.
The unparticle is a scale invariant stuff in a hidden sector.
Its interactions with the SM particles are well described by an effective
theory formalism.
\par
The most striking feature of the unparticle is its unusual phase space with
non-integral scaling dimension $d_\calU$.
For an unparticle operator of scaling dimension $d_\calU$,
the unparticle appears as a non-integral number $d_\calU$ of invisible
massless particles.
After the Georgi's suggestion, there have been a lot of phenomenological
studies on unparticles \cite{Cheung,Yuan,Fox,DEQ,Stephanov,jplee}.
\par
Among other couplings between SM fields and unparticles, Higgs-unparticle
interaction is very interesting because its coupling,
$\lambda_{\Phi\Phi\calU}(\Phi^\dagger\Phi)\calO_\calU$, 
 is relevant \cite{Fox}.
Here $\Phi$ is a fundamental Higgs, $\calO_\calU$ is a scalar unparticle
operator with scaling dimension $1<d_\calU<2$, and
$\lambda_{\Phi\Phi\calU}$ is the coupling constant.
\par
This work is motivated by the observation that the scalar unparticle
operator $\calO_\calU$ is {\em equivalent} to $d_\calU$ number of massless
particles \cite{jplee2}.
This feature of unparticle operator coupled with Higgs sector is already pointed
out in the literature \cite{See}.
We propose a new type of scalar potential
\begin{equation}
V_{int}\sim\lambda(\Phi^\dagger\Phi)(\phi^*\phi)^{d_\calU/2}~,
\end{equation}
where $\phi$ is a massless scalar field with $[\phi]=1$.
\par
When one considers the scalar potential containing the form of
$V_{int}$, it inevitably introduces a mass scale through the dimensionful
coupling.
One may expect that there is a nontrivial minimum along the radial direction
{\em at tree level} for $V\supset V_{int}$.
The main result of this work is that this is indeed the case.
In other words, interactions between the SM fields and unparticle sector
themselves break the electroweak symmetry.
\par
After the EWSB occurs one expands the scalar fields around the vacuum.
The resulting fluctuations mix up with each other to form two physical scalar
states.
In this simple setup, it is quite natural to identify a heavy state as Higgs.
The other light state has a mass proportional to $(2-d_\calU)$ which vanishes
as $d_\calU\to 2$.
This is the remnant of the fact that $V_0$ has a massless scalar at tree level
as a pseudo Goldstone boson from the conformal symmetry breaking.
The unparticle sector thus no longer remains scale-invariant after the EWSB.
So the interaction $V_{int}$ induces {\em both} EWSB in the SM sector and the
scale-invariance breaking in the unparticle sector.
We find that all of these things can happen for acceptable values of the
parameters of this setup.
\par
In the earlier works of \cite{DEQ}, the EWSB with unparticles was considered
in the context of deconstruction \cite{Stephanov}.
But the scalar potential of \cite{DEQ} contains only the usual polynomials of
the fields.
The unusual scaling behavior of the unparticle was encoded in the deconstructed
version of the unparticle decay constant,
while in this work it is simplified with the fractional power of the fields.
We find that the EWSB conditions on the parameters here are rather stronger,
and the predicted new scalar particle is always lighter than the Higgs boson.
\par
Recently it was proposed that the unparticles are gauged to become a Higgs 
\cite{Terning}.
This "Unhiggs" can also break EWS and unitarize $WW$ scattering.
\section{Scalar potential and the mass spectrum}
The proposed scalar potential has the form of
\begin{eqnarray}
V(\Phi,\phi)&=&\lambda_0(\Phi^\dagger\Phi)^2+\lambda_1(\phi^*\phi)^2\\\nn
 && +2\lambda_2\mu^{2-d_\calU}(\Phi^\dagger\Phi)
 (\phi^*\phi)^{{d_\calU}/{2}}~,
\label{V}
\end{eqnarray}
where $\lambda_0$ is assumed to be positive. Here the mass
dimension of $\phi$ is $1$ and a dimension-1 parameter $\mu$ is
inserted to make $\lambda_2$ dimensionless. 
The minimum of $V$ lies along some ray $\Phi_i=\rho N_i$ \cite{GW},
where $\vec{N}$ is a unit vector in the field space $\Phi_i=(\Phi,\phi)$. 
In unitary gauge, the fields are parameterized as
\begin{equation}
\Phi=\frac{\rho}{\sqrt{2}}
\left(\begin{array}{c}
0\\N_0\end{array}\right)~,~~~
\phi=\frac{\rho}{\sqrt{2}}N_1~,
\end{equation}
where $N_0^2+N_1^2=1$.
The scalar potential becomes
\begin{equation}
V(\rho,{\vec N})=\frac{\rho^4}{4}\left[
\lambda_0N_0^4+\lambda_1N_1^4
+\left(\frac{\rhohat^2}{2}\right)^{-\epsilon}
 2\lambda_2N_0^2N_1^{d_\calU}\right]~,
\end{equation}
where $d_\calU\equiv 2-2\epsilon$, and $\rhohat\equiv\rho/\mu$.
\par
The stationary condition for $V$ along the $\vec N$ direction for some
specific unit vector ${\vec N}={\vec n}$,
$(\partial V/\partial N_i)_{\vec n}=0$, 
combined with the normalization of $\vec n$ ($n_1^2+n_2^2=1$), gives
\begin{equation}
\label{rho}
n_0^2=\frac{\sqrt{2\lambda_1}}
 {\sqrt{d_\calU\lambda_0}+\sqrt{2\lambda_1}}~,~~~
n_1^2=\frac{\sqrt{d_\calU\lambda_0}}
 {\sqrt{d_\calU\lambda_0}+\sqrt{2\lambda_1}}~.
\end{equation}
One can easily find that the minimum of $V(\rho,{\vec n})$ occurs at
\begin{equation}
\rho=\rho_0\equiv\Bigg(
 -\frac{2^\epsilon\lambda_2 n_1^{d_\calU}}{\lambda_0n_0^2}
 \Bigg)^{\frac{1}{2\epsilon}}\mu~.
\label{rho0}
\end{equation}
\par
It should be noted that when $d_\calU\to2$, $\rho_0$ goes to $0$ or infinity
depending on the values of $\lambda_{0,2}$ and $n_{0,1}$.
Since the vacuum expectation value of $\rho$ is directly proportional to the
mass scale of the theory (e.g., gauge boson masses, Higgs masses, etc.),
it is not desirable if $\rho_0$ gets too small or too large for $d_\calU\to2$.
We require that $\rho_0$ is stable for $d_\calU\to2$ ($\epsilon\to0$).
A little algebra shows that this requirement is satisfied if
\begin{equation}
\lambda_2=-\sqrt{\lambda_0\lambda_1}\equiv\lambdabar~.
\end{equation}
\par
When $\lambda_{1,2}$ are turned on, the potential $V$ develops the VEV $v$
at $\rho=\rho_0$ and the fields $\Phi$ and $\phi$ get fluctuations $h$ and
$s$ around $v$.
The scalar potential now becomes
\begin{eqnarray}
V(h,s)&=&
\frac{\lambda_0}{4}(n_0\rho_0+h)^4+\frac{\lambda_1}{4}(n_1\rho_0+s)^4\\\nn
&&+2^{-d_\calU/2}\lambda_2\mu^{2\epsilon}
(n_0\rho_0+h)^2(n_1\rho_0+s)^{d_\calU}~.
\end{eqnarray}
The mass squared matrix 
\begin{equation}
(M^2)_{i,j}=\frac{\partial^2V}{\partial\psi_i\partial\psi_j}\Bigg|_0~,
\end{equation}
where $\psi_i=(h,s)$, gives
two eigenvalues of $M^2$ corresponding to the heavy and light scalar
mass squared respectively:
\begin{eqnarray}
m_{h,\ell}^2&=& \frac{\rho_0^2\sqrt{2\lambda_0\lambda_1}}
{\sqrt{d_\calU\lambda_0}+\sqrt{2\lambda_1}}\Bigg\{
\sqrt{\lambda_0}\\\nn
&&+\left(2-\frac{d_\calU}{2}\right)\sqrt{\frac{d_\calU}{2}\lambda_1}
\pm\sqrt{D}\Bigg\}~,
\end{eqnarray}
where
\begin{equation}
D=\lambda_0+\left(2-\frac{d_\calU}{2}\right)^2\frac{d_\calU}{2}\lambda_1
+\left(\frac{3d_\calU}{2}-2\right)\sqrt{2d_\calU\lambda_0\lambda_1}~.
\end{equation}
\par
When $d_\calU=2$, the light scalar is massless at tree level.
The reason is that it corresponds to the pseudo Goldstone boson from
the spontaneous symmetry breaking of the conformal symmetry
\cite{GW,Goldberger}.
\par
But for $\epsilon=1-d_\calU/2\ll 1$, we have found that
$m_\ell^2/m_h^2\sim\epsilon$ at tree level. Thus the new light
scalar and Higgs boson masses are good probes to the hidden
unparticle sector.
\par
The vacuum expectation value $\rho_0$ is related to the gauge
boson ($W$) masses as
\begin{equation}
(n_0\rho_0)^2=\frac{1}{\sqrt{2}G_F}=(246~{\rm GeV})^2\equiv
v_0^2~.
\end{equation}
From the value of $\rho_0$ and $\lambda_2$, one has
\begin{equation}
\vhat_0^2=2\left(\frac{d_\calU}{2}\right)^{\frac{d_\calU}{2(2-d_\calU)}}
\sqrt{\frac{\lambda_1}{\lambda_0}}~.
\label{vhat0}
\end{equation}
where $\vhat_0=v_0/\mu$.
The right-hand-side of Eq.\ (\ref{vhat0}) is a slow varying function of
$d_\calU$.
If one chooses $\mu=v_0$, the ratios of couplings are
\begin{eqnarray}
\frac{\lambda_1}{\lambda_0}
&=&\frac{1}{4}\left(\frac{2}{d_\calU}\right)^{\frac{d_\calU}{2-d_\calU}}
\longrightarrow\frac{e}{4}\simeq 0.68~~{\rm as}~d_\calU\to 2~, \nn\\
\frac{\lambda_2}{\lambda_0}&=&-\sqrt{\frac{\lambda_1}{\lambda_0}}
\longrightarrow-0.82~.
\label{fixratio}
\end{eqnarray}
When $d_\calU=1$, $\lambda_1/\lambda_0=0.5$ and
$\lambda_2/\lambda_0\simeq-0.71$.
Since the ratios are of order 1 for all range over $d_\calU$,
the scale of $\mu$ around the weak scale is a reasonable choice.
\begin{figure}
\includegraphics{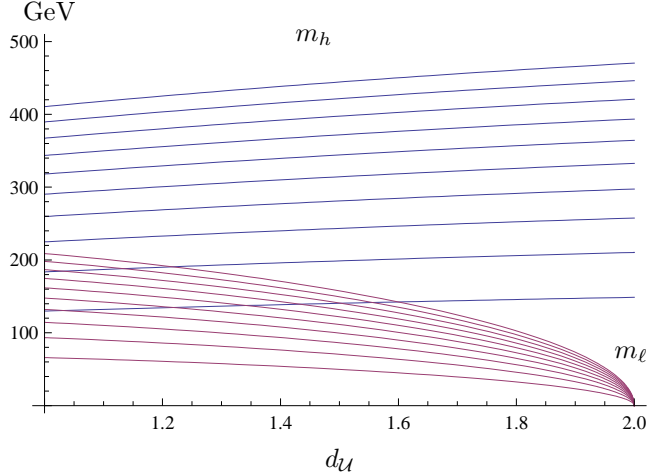}
\caption{\label{mass}Scalar masses $m_h$ and $m_\ell$ as a function of $d_\calU$
for $\lambda_0=$ 0.1, 0.2, $\cdots$, 1.0, from bottom to top.}
\end{figure}
As given in Fig.\ \ref{mass}, $m_h$ is rather inert with respect to
$d_\calU$ while $m_\ell$ is not.
With the condition of Eq.\ (\ref{fixratio}), both $m_{h,\ell}$ are proportional
to $\sim\sqrt{\lambda_0}$.
One can find that
\begin{eqnarray}
130(149)~{\rm GeV}&\lesssim& m_h\lesssim 411(470)~{\rm GeV}~,\nn\\
66~{\rm GeV}&\lesssim& m_\ell\lesssim 209~{\rm GeV}~,
\label{massrange}
\end{eqnarray}
for $d_\calU=1(2)$.
If the scalar masses turned out to be quite different from
Eq.\ (\ref{massrange}), then the value of $\mu$ should be rearranged to fit
the data.
But in this case one would have to explain why that value of $\mu$ is so
different from $v_0$, the electroweak scale.

\section{Conclusion}
In this talk we suggest a new scalar potential with a fractional power of
fields from hidden sector inspired by the scalar unparticle operator.
Unlike the usual potential of marginal coupling, the new one develops VEV at
tree level.
In this picture, the EWSB occurs when the unparticle sector begins to interact
with the SM sector.
When the scaling dimension $d_\calU$ departs from the value of $2$
a new scale (of the order of $\sim 1/\sqrt{G_F}$) is introduced in the scalar
potential through the relevant coupling,
and the electroweak symmetry is broken at tree level.
In other words, the EWSB occurs when the hidden sector enters the regime of
scale invariance, i.e., unparticles.
In view of the unparticle sector, the new potential also breaks the scale
invariance of the hidden sector.
\par
Once the electroweak symmetry is broken, the scalar fields from SM and hidden
sector mix together to form two massive physical states.
The heavy one is identified as Higgs, while the light one is a new particle
of mass around $\lesssim 210$ GeV.
It might be that the latter be soon discovered at the LHC.


\begin{thebibliography}{999}
\bibitem{Meissner}
  K.~A.~Meissner and H.~Nicolai,
  Phys.\ Lett.\  B {\bf 648}, 312 (2007).
\bibitem{Espinosa}
  J.~R.~Espinosa and M.~Quiros,
  Phys.\ Rev.\  D {\bf 76}, 076004 (2007).
\bibitem{CNW}
W.~F.~Chang, J.~N.~Ng and J.~M.~S.~Wu,
  Phys.\ Rev.\  D {\bf 75}, 115016 (2007).
\bibitem{FKV}
  R.~Foot, A.~Kobakhidze and R.~R.~Volkas,
  Phys.\ Lett.\  B {\bf 655}, 156 (2007).
\bibitem{CW}
  S.~R.~Coleman and E.~Weinberg,
  Phys.\ Rev.\  D {\bf 7}, 1888 (1973).
\bibitem{Georgi}
  H.~Georgi,
  Phys.\ Rev.\ Lett.\  {\bf 98}, 221601 (2007);
Phys.\ Lett.\  B {\bf 650}, 275 (2007);
\bibitem{Cheung}
 K.~Cheung, W.~Y.~Keung and T.~C.~Yuan,
  Phys.\ Rev.\ Lett.\  {\bf 99}, 051803 (2007),
  Phys.\ Rev.\  D {\bf 76}, 055003 (2007);
M.~Luo and G.~Zhu,
Phys.\ Lett.\  B {\bf 659}, 341 (2008);
Y.~Liao,
  Phys.\ Rev.\  D {\bf 76}, 056006 (2007);
T.~Kikuchi and N.~Okada,
  arXiv:0707.0893 [hep-ph];
A.~Lenz,
  Phys.\ Rev.\  D {\bf 76}, 065006 (2007);
J.~R.~Mureika,
  Phys.\ Lett.\  B {\bf 660}, 561 (2008);
B.~Grinstein, K.~Intriligator and I.~Z.~Rothstein,
  arXiv:0801.1140 [hep-ph].
\bibitem{Yuan}
K.~Cheung, W.~Y.~Keung and T.~C.~Yuan,
this proceeding;
  arXiv:0809.0995 [hep-ph].
\bibitem{Fox}
P.~J.~Fox, A.~Rajaraman and Y.~Shirman,
  Phys.\ Rev.\  D {\bf 76}, 075004 (2007).
\bibitem{DEQ}
A.~Delgado, J.~R.~Espinosa and M.~Quiros,
  JHEP {\bf 0710}, 094 (2007).
\bibitem{Stephanov}
  M.~A.~Stephanov,
  Phys.\ Rev.\  D {\bf 76}, 035008 (2007).
\bibitem{jplee}
J.~P.~Lee,
  arXiv:0710.2797 [hep-ph].
\bibitem{jplee2}
J.~P.~Lee,
  arXiv:0803.0833 [hep-ph].
\bibitem{See}
See, for example, Ref.\ \cite{DEQ}.
\bibitem{Terning}
D.~Stancato and J.~Terning,
  arXiv:0807.3961 [hep-ph].
\bibitem{GW}
  E.~Gildener and S.~Weinberg,
  Phys.\ Rev.\  D {\bf 13}, 3333 (1976).
\bibitem{Goldberger}
  W.~D.~Goldberger, B.~Grinstein and W.~Skiba,
  arXiv:0708.1463 [hep-ph].
\end{thebibliography}
\end{document}